\documentstyle[12pt]{article}
\newcommand{\nc}{\newcommand}
\nc{\renc}{\renewcommand}

\nc{\half}{{\textstyle{1\over2}}}

\nc{\etal}{\mbox{\it et al. }}
\nc{\ie}{{\it i.e.}}
\nc{\eg}{{\it e.g.}}

\renc{\thefootnote}{\arabic{footnote}}
\nc{\capt}[1]{{\bf Figure.} {\small\sl #1}}

\nc{\eqs}[2]{\mbox{Eqs.~(\ref{#1},\,\ref{#2})}}
\nc{\eq}[1]{\mbox{Eq.~(\ref{#1})}}

\nc{\figs}[2]{\mbox{Figs.~(\ref{#1},\,\ref{#2})}}
\nc{\fig}[1]{\mbox{Fig~.(\ref{#1})}}

\nc{\tag}[1]{\label{#1} \marginpar{{\footnotesize #1}}}
\nc{\mtag}[1]{\label{#1} \mbox{\marginpar{{\footnotesize #1}}}}
\renc{\baselinestretch}{1.2}
\newlength{\overeqskip}
\newlength{\undereqskip}
\nc{\be}[1]{\begin{equation} \mbox{$\label{#1}$}}
\nc{\bea}[1]{\begin{eqnarray} \mbox{$\label{#1}$}}
\nc{\Section}[2]{\section{#2}\label{#1}}
\nc{\Bibitem}[1]{\bibitem{#1}}
\nc{\Label}[1]{\label{#1}}
\nc{\eea}{\vspace{\undereqskip}\end{eqnarray}}
\nc{\ee}{\vspace{\undereqskip}\end{equation}}
\nc{\bdm}{\begin{displaymath}}
\nc{\edm}{\end{displaymath}}
\nc{\dpsty}{\displaystyle}
\nc{\bc}{\begin{center}}
\nc{\ec}{\end{center}}
\nc{\ba}{\begin{array}}
\nc{\ea}{\end{array}}
\nc{\bab}{\begin{abstract}}
\nc{\eab}{\end{abstract}}
\nc{\btab}{\begin{tabular}}
\nc{\etab}{\end{tabular}}
\nc{\bit}{\begin{itemize}}
\nc{\eit}{\end{itemize}}
\nc{\ben}{\begin{enumerate}}
\nc{\een}{\end{enumerate}}
\nc{\bfig}{\begin{figure}}
\nc{\efig}{\end{figure}}
\nc{\arreq}{&\!=\!&}
\nc{\arrmi}{&\!-\!&}
\nc{\arrpl}{&\!+\!&}
\nc{\arrap}{&\!\!\!\approx\!\!\!&}
\nc{\non}{\nonumber\\*}
\nc{\align}{\!\!\!\!\!\!\!\!&&}

\def\lsim{\; \raise0.3ex\hbox{$<$\kern-0.75em
      \raise-1.1ex\hbox{$\sim$}}\; }
\def\gsim{\; \raise0.3ex\hbox{$>$\kern-0.75em
      \raise-1.1ex\hbox{$\sim$}}\; }
\nc{\DOT}{\hspace{-0.08in}{\bf .}\hspace{0.1in}}
\nc{\Laada}{\hbox {$\sqcap$ \kern -1em $\sqcup$}}
\nc\loota{{\scriptstyle\sqcap\kern-0.55em\hbox{$\scriptstyle\sqcup$}}}
\nc\Loota{{\sqcap\kern-0.65em\hbox{$\sqcup$}}}
\nc\laada{\Loota}
\nc{\qed}{\hskip 3em \hbox{\BOX} \vskip 2ex}

\nc{\real}{{\rm I \! R}}
\nc{\Z}{{\sf Z \!\!\! Z}}
\nc{\complex}{{\rm C\!\!\! {\sf I}\,\,}}
\def\bigid{\leavevmode\hbox{\small1\kern-3.8pt\normalsize1}}
\def\id{\leavevmode\hbox{\small1\kern-3.3pt\normalsize1}}
\nc{\slask}{\!\!\!/}
\nc{\bis}{{\prime\prime}}
\nc{\pa}{\partial}
\nc{\na}{\nabla}
\nc{\ra}{\rangle}
\nc{\la}{\langle}
\nc{\goto}{\rightarrow}
\nc{\swap}{\leftrightarrow}

\nc{\EE}[1]{ \mbox{$\cdot10^{#1}$} }
\nc{\abs}[1]{\left|#1\right|}
\nc{\at}[2]{\left.#1\right|_{#2}}
\nc{\norm}[1]{\|#1\|}
\nc{\abscut}[2]{\Abs{#1}_{\scriptscriptstyle#2}}
\nc{\vek}[1]{{\rm\bf #1}}
\nc{\integral}[2]{\int\limits_{#1}^{#2}}
\nc{\inv}[1]{\frac{1}{#1}}
\nc{\dd}[2]{{{\partial #1}\over{\partial #2}}}
\nc{\ddd}[2]{{{{\partial}^2 #1}\over{\partial {#2}^2}}}
\nc{\dddd}[3]{{{{\partial}^2 #1}\over
        {\partial #2 \partial #3}}}
\nc{\dder}[2]{{{d #1}\over{d #2}}}
\nc{\ddder}[2]{{{d^2 #1}\over{d {#2}^2}}}
\nc{\dddder}[3]{{d^2 #1}\over
        {d #2 d #3}}
\nc{\dx}[1]{d\,^{#1}x}
\nc{\dy}[1]{d\,^{#1}y}
\nc{\dz}[1]{d\,^{#1}z}
\nc{\dl}[1]{\frac{d\,^{#1}l}{(2\pi)^{#1}}}
\nc{\dk}[1]{\frac{d\,^{#1}k}{(2\pi)^{#1}}}
\nc{\dq}[1]{\frac{d\,^{#1}q}{(2\pi)^{#1}}}

\nc{\cc}{\mbox{$c.c.$ }}
\nc{\hc}{\mbox{$h.c.$ }}
\nc{\cf}{cf.\ }
\nc{\erfc}{{\rm erfc}}
\nc{\Tr}{{\rm Tr\,}}
\nc{\tr}{{\rm tr\,}}
\nc{\pol}{{\rm pol}}
\nc{\sign}{{\rm sign}}
\nc{\bfT}{{\bf T }}

\def\GeV{{\rm\ GeV}}
\def\MeV{{\rm\ MeV}}

\nc{\cA}{{\cal A}}
\nc{\cB}{{\cal B}}
\nc{\cD}{{\cal D}}
\nc{\cE}{{\cal E}}
\nc{\cG}{{\cal G}}
\nc{\cH}{{\cal H}}
\nc{\cL}{{\cal L}}
\nc{\cO}{{\cal O}}
\nc{\cT}{{\cal T}}
\nc{\cN}{{\cal N}}
\nc{\rvac}[1]{|{\cal O}#1\rangle}
\nc{\lvac}[1]{\langle{\cal O}#1|}
\nc{\rvacb}[1]{|{\cal O}_\beta #1\rangle}
\nc{\lvacb}[1]{\langle{\cal O}_\beta #1 |}
\nc{\bb}{\bar{\beta}}
\nc{\bt}{\tilde{\beta}}
\nc{\ctH}{\tilde{\cal H}}
\nc{\chH}{\hat{\cal H}}
\nc{\1}{\aa}
\nc{\2}{\"{a}}
\nc{\3}{\"{o}}
\nc{\4}{\AA}
\nc{\5}{\"{A}}
\nc{\6}{\"{O}}
\nc{\al}{\alpha}
\nc{\g}{\gamma}
\nc{\Del}{\Delta}
\nc{\e}{\epsilon}
\nc{\eps}{\epsilon}
\nc{\lam}{\lambda}
\nc{\om}{\omega}
\nc{\Om}{\Omega}
\nc{\ve}{\varepsilon}
\nc{\mn}{{\mu\nu}}
\nc{\k}{\kappa}
\nc{\vp}{\varphi}

\nc{\advp}[3]{{\it  Adv.\ in\ Phys.\ }{{\bf #1} {(#2)} {#3}}}
\nc{\annp}[3]{{\it  Ann.\ Phys.\ (N.Y.)\ }{{\bf #1} {(#2)} {#3}}}
\nc{\apl}[3]{{\it  Appl. Phys. Lett. }{{\bf #1} {(#2)} {#3}}}
\nc{\apj}[3]{{\it  Ap.\ J.\ }{{\bf #1} {(#2)} {#3}}}
\nc{\apjl}[3]{{\it  Ap.\ J.\ Lett.\ }{{\bf #1} {(#2)} {#3}}}
\nc{\app}[3]{{\it Astropart.\ Phys.\ }{{\bf #1} {(#2)} {#3}}}
\nc{\cmp}[3]{{\it  Comm.\ Math.\ Phys.\ }{{ \bf #1} {(#2)} {#3}}}
\nc{\cqg}[3]{{\it  Class.\ Quant.\ Grav.\ }{{\bf #1} {(#2)} {#3}}}
\nc{\epl}[3]{{\it  Europhys.\ Lett.\ }{{\bf #1} {(#2)} {#3}}}
\nc{\ijmp}[3]{{\it Int.\ J.\ Mod.\ Phys.\ }{{\bf #1} {(#2)} {#3}}}
\nc{\ijtp}[3]{{\it Int.\ J.\ Theor.\ Phys.\ }{{\bf #1} {(#2)} {#3}}}
\nc{\jmp}[3]{{\it  J.\ Math.\ Phys.\ }{{ \bf #1} {(#2)} {#3}}}
\nc{\jpa}[3]{{\it  J.\ Phys.\ A\ }{{\bf #1} {(#2)} {#3}}}
\nc{\jpc}[3]{{\it  J.\ Phys.\ C\ }{{\bf #1} {(#2)} {#3}}}
\nc{\jap}[3]{{\it J.\ Appl.\ Phys.\ }{{\bf #1} {(#2)} {#3}}}
\nc{\jpsj}[3]{{\it J.\ Phys.\ Soc.\ Japan\ }{{\bf #1} {(#2)} {#3}}}
\nc{\lmp}[3]{{\it Lett.\ Math.\ Phys.\ }{{\bf #1} {(#2)} {#3}}}
\nc{\mpl}[3]{{\it  Mod.\ Phys.\ Lett.\ }{{\bf #1} {(#2)} {#3}}}
\nc{\ncim}[3]{{\it  Nuov.\ Cim.\ }{{\bf #1} {(#2)} {#3}}}
\nc{\np}[3]{{\it  Nucl.\ Phys.\ }{{\bf #1} {(#2)} {#3}}}
\nc{\pr}[3]{{\it Phys.\ Rev.\ }{{\bf #1} {(#2)} {#3}}}
\nc{\pra}[3]{{\it  Phys.\ Rev.\ A\ }{{\bf #1} {(#2)} {#3}}}
\nc{\prb}[3]{{\it  Phys.\ Rev.\ B\ }{{{\bf #1} {(#2)} {#3}}}}
\nc{\prc}[3]{{\it  Phys.\ Rev.\ C\ }{{\bf #1} {(#2)} {#3}}}
\nc{\prd}[3]{{\it  Phys.\ Rev.\ D\ }{{\bf #1} {(#2)} {#3}}}
\nc{\prl}[3]{{\it Phys.\ Rev.\ Lett.\ }{{\bf #1} {(#2)} {#3}}}
\nc{\pl}[3]{{\it  Phys.\ Lett.\ }{{\bf #1} {(#2)} {#3}}}
\nc{\prep}[3]{{\it Phys\. Rep.\ }{{\bf #1} {(#2)} {#3}}}
\nc{\prsl}[3]{{\it Proc.\ R.\ Soc.\ London\ }{{\bf #1} {(#2)} {#3}}}
\nc{\ptp}[3]{{\it  Prog.\ Theor.\ Phys.\ }{{\bf #1} {(#2)} {#3}}}
\nc{\ptps}[3]{{\it  Prog\ Theor.\ Phys.\ suppl.\ }{{\bf #1} {(#2)} {#3}}}
\nc{\physa}[3]{{\it  Physica\ A\ }{{\bf #1} {(#2)} {#3}}}
\nc{\physb}[3]{{\it  Physica\ B\ }{{\bf #1} {(#2)} {#3}}}
\nc{\phys}[3]{{\it Physica\ }{{\bf #1} {(#2)} {#3}}}
\nc{\rmp}[3]{{\it  Rev.\ Mod.\ Phys.\ }{{\bf #1} {(#2)} {#3}}}
\nc{\rpp}[3]{{\it Rep.\ Prog.\ Phys.\ }{{\bf #1} {(#2)} {#3}}}
\nc{\sjnp}[3]{{\it Sov.\ J.\ Nucl.\ Phys.\ }{{\bf #1} {(#2)} {#3}}}
\nc{\spjetp}[3]{{\it Sov.\ Phys.\ JETP\ }{{\bf #1} {(#2)} {#3}}}
\nc{\yf}[3]{{\it Yad.\ Fiz.\ }{{\bf #1} {(#2)} {#3}}}
\nc{\zetp}[3]{{\it Zh.\ Eksp.\ Teor.\ Fiz.\  }{{\bf #1}  {(#2)} {#3}}}
\nc{\zp}[3]{{\it Z.\ Phys.\ }{{\bf #1} {(#2)} {#3}}}
\nc{\ibid}[3]{{\sl ibid.\ }{{\bf #1} {#2} {#3}}}
\nc{\rf}[1]{(\ref{#1})}
\nc{\nn}{\nonumber \\*}
\nc{\bfB}{\bf{B}}
\nc{\bfv}{\bf{v}}
\nc{\bfx}{\bf{x}}
\nc{\bfy}{\bf{y}}
\nc{\vx}{\vec{x}}
\nc{\vy}{\vec{y}}
\nc{\oB}{\overline{B}}
\nc{\oI}{\overline{I}}
\nc{\oR}{\overline{R}}
\nc{\rar}{\rightarrow}
\nc{\ti}{\times}
\nc{\slsh}{\hskip-5pt/}
\nc{\sm}{Standard~Model~}
\nc{\MP}{M_{\rm Pl}}
\nc{\tp}{t_{\rm Pl}}
\nc{\ave}{\bar{E}}

\nc{\eff}{{\rm eff}}
\nc{\kk}{\vek{k}}
\nc{\pp}{{\rm p}}
\nc{\ga}{g_{a\gamma}}
\nc{\vv}{\\}
\nc{\eee}{{\bf E}}
\nc{\bbb}{{\bf B}}
\nc{\qcd}{T_{\rm QCD}}
\nc{\G}{\rm \ G}
\def\vec#1{{\bf #1}}
\begin{document}

{\title{\vskip-2truecm{\hfill {{\small \\
        % \hfill hep-ph/9801434\\
        }}\vskip 1truecm}
{\bf Transport coefficients in the early universe}}

%\vspace{1.2cm}

{\author{
{\sc Jarkko Ahonen$^{1}$}\\
{\sl\small Department of Physics, P.O. Box 9,
FIN-00014 University of Helsinki,
Finland}
}
\maketitle
\vspace{2cm}
%\newpage
\begin{abstract}
\noindent
We calculate numerically the electrical conductivity $\sigma$,  
heat conductivity $\kappa$ and shear viscosity $\eta$
of the hot plasma present in the early universe for the
temperature interval $1\MeV\lsim T\lsim
10\GeV$. We use the Boltzmann collision equation to
compute all the scattering matrix elements and regulate them by the
thermal masses of the $t$- and $u$-channel particles.
No leading order approximation is needed because of
the numerical integration routines used. 
\end{abstract}
\vfil
\footnoterule
{\small $^1$jtahonen@pcu.helsinki.fi}
\thispagestyle{empty}
\newpage
\setcounter{page}{1}

%%%%%%%%%%%%%%%%%%%%%%%%%%%%%%%%%%%%%%%%%%%%%%%%%%%%%%%%%%%%%%%%%%%%%%%%%%%%%%
\section{Introduction}
%%%%%%%%%%%%%%%%%%%%%%%%%%%%%%%%%%%%%%%%%%%%%%%%%%%%%%%%%%%%%%%%%%%%%%%%%%%%%%
The transport properties of the hot plasma present in the early universe are
of great interest. The transport coefficients play a significant role in the
phase transitions of the early universe \cite{18,19,20,21}, the creation and
development of the primordial magnetic fields \cite{15,16} and finally in
the creation of the primordial density perturbations and therefore in the 
galaxy formation 
\cite{13,14}. 

There has been many attempts to estimate the transport
coefficients. 
(Textbook estimates for thermal conductivity, shear and
bulk viscosity in 
the ultrarelativistic plasma were given in \cite{10} and \cite{11}.)
In \cite{7} the viscosities of a pure gluon plasma and of a
quark-gluon plasma were computed in the weak coupling limit from a 
variational solution
to the Boltzmann equation. In \cite{8} the transport coefficients 
were calculated for plasmas
interacting through strong, electromagnetic and weak interactions to leading
order in the interaction strength. The rates
of momentum and thermal relaxation, electrical conductivity and viscosities of
quark-gluon and electrodynamic plasmas were also included in \cite{8}. 
In \cite{9} the transport coefficients
and relaxation times were calculated to leading orders in the coupling
constant for degenerate quark matter within perturbative QCD for 
temperatures and inverse screening lengths much smaller than the quark 
chemical potential. The viscosities of the QCD-plasma were considered in
\cite{24}. Textbook estimates for thermal conductivity and 
viscosity in the 
ultrarelativistic plasma were given in \cite{10} and \cite{11}. 
The conductivity
of a relativistic plasma has also been considered in \cite{12} by reformulating
the collision operator for a relativistic plasma in terms of an expansion in
spherical harmonics.

However, all the previous estimates involve approximations.
The calculations have been performed only to the leading (logarithmic) orders
and certain scattering reactions 
present in the hot primordial plasma have been neglected. In the present 
paper, the only approximation we make is to assume 
that when particles appear in the heat bath of the primordial plasma, their 
thermal velocities are high enough to allow us to treat them 
ultrarelativistic and therefore massless. We consider radiation dominated
plasma with temperatures from about 1 MeV to about 10 GeV.

The viscous damping and heat conducting effects affect the first order
phase transitions in the early universe. During such phase transitions, 
instabilities may occur when the transport of latent heat is dominated by 
the fluid flow. In \cite{18} this was studied in the EW transition in the small
velocity limit, in \cite{19} in the QCD transition, in \cite{20} for 
cosmological detonation fronts and in \cite{21} for general first order 
transitions with either large or small bubble wall velocities. The 
instabilities can be damped by finite viscosity and heat conductivity due to
the diffusion of radiation on small length scales. 

When considering the creation and development of primordial magnetic fields it 
is also of importance to know the transport coefficients of the surrounding 
primordial plasma \cite{15,16}. Finite electrical conductivity leads 
to the
diffusion of magnetic fields and therefore is of importance when trying to
explain the further evolution of an initial seed magnetic field (see e.g. 
\cite{17}). The instabilities in the first order phase transitions can
be shown to create seed magnetic fields \cite{16}. Therefore the seed 
primordial magnetic fields also depend on the size of the transport 
coefficients in the primordial plasma in a crucial way. In particular, the 
seed fields created in the QCD phase
transition are highly dependent on the neutrino viscosity \cite{16}.

Galaxy formation \cite{13} is likewise affected by the thermal coefficients 
in the primordial plasma \cite{14}. Viscosity tends to heat up the 
plasma, and on the other
hand thermal conduction transfers heat from regions of high temperature to
regions at lower temperature. The effects affect the structure
formation and thus in order to make precise models of the galaxy formation
it is important to know the values of the transport coefficients.

It should be noted that two major effects are involved when heat transport
and viscosity are studied. These effects are the number of possible 
scattering reactions, i.e. the number of particles present in the plasma,
and the Debye mass present in the propagator. As we will discuss in Section 4,
the first effect will decrease the transport coefficients while the second
effect will increase them. The net behaviour of the coefficients
depend on the exact interplay between these two effects.

The different interactions give contributions of different strengths to the
electrical conductivity as was shown in \cite{1}. However, in \cite{1}
only strong and electromagnetic interactions were considered.
Here we also deal with the scattering reactions involving neutrinos.

The paper is organized as follows.
In section 2 we introduce the reader to the method of calculation and, 
using the full definition of the energy-stress tensor $T^{\mu\nu}$, we define
the transport coefficients presented here.
In section 3 we reconsider electrical conductivity.
We first discuss what is the proper way to introduce the Debye screening to the
propagators in the matrix elements. We then include additional scattering 
reactions not present in previous work and
present a new value for the electrical conductivity. We also compare our
results with other recent estimates for the electrical conductivity in the 
hot plasma. In Section 4 and 5 we apply our technique to calculating 
thermal conductivity
and shear viscosity in the hot plasma, respectively, and compare our results 
with other recent estimates. Section 6 contains a summary of the results
and a brief discussion on their relevance for the early universe.
%%%%%%%%%%%%%%%%%%%%%%%%%%%%%%%%%%%%%%%%%%%%%%%%%%%%%%%%%%%%%%%%%%%%%%%%%%%%%%
\section{The general formalism}
%%%%%%%%%%%%%%%%%%%%%%%%%%%%%%%%%%%%%%%%%%%%%%%%%%%%%%%%%%%%%%%%%%%%%%%%%%%%%%
\subsection{The energy-stress tensor}
We start by defining the transport coefficients. We do this by using the
energy-stress tensor $T^{\mu\nu}$:
\be{stress}
T^{\mu\nu}=(\rho +p)U^\mu U^\nu -pg^{\mu\nu}+T_{\mbox{{\tiny TP}}}^{\mu\nu}+
T_{\mbox{{\tiny EM}}}^{\mu\nu},
\ee
where $\rho$ is the total energy density, $p$ the pressure, $U^{\mu}$ the
velocity four-vector ($U^\mu U_\mu =-1$) and $g^{\mu\nu}$ the metric tensor. 
$T_{\mbox{{\tiny EM}}}^{\mu\nu}$ is the usual electromagnetic
energy-stress tensor (see e.g. \cite{22}) and $T_{\mbox{{\tiny TP}}}^{\mu\nu}$
represents the non-ideal contributions from a finite viscosity
and heat conductivity caused by the diffusion of the various particles present
in the plasma.

To find out the meaning of the transport coefficients we now follow \cite{10} 
and derive the form of $T_{\mbox{{\tiny TP}}}^{\mu\nu}$. 
Because of dissipation, we must also add a correction term to the particle
current $N^{\mu}$: 
\be{number}
N^{\mu}=nU^{\mu} + N_{d}^{\mu},
\ee
where $n$ is the particle number density and $N_{d}^{\mu}$ the correction term.
To avoid the ambiguities generated by the adding of extra terms induced by 
dissipation and diffusion, we should define what is meant by the number
and energy densities:
\bea{defner}
n&\equiv &-U^\mu N_\mu \ , \cr
\rho&\equiv &U_\alpha U_\beta T^{\alpha\beta} \ , \cr
U^\mu &\equiv &(-N^\nu N_\nu )^{-1/2}N^\mu .
\eea
Inserting the definitions \eq{defner} into \eq{stress} and \eq{number} we
can easily see that the following conditions must be satisfied:
\bea{cond}
N_{d}^{\mu}&=&0 \ , \cr
\null
U_\alpha U_\beta T^{\alpha\beta}_{TP}&=&0 \ .
\eea
Using the conservation laws for the fluid:
\bea{fluidcons}
\partial_\nu T^{\mu\nu}&=&0 \ , \cr
\null
\partial_\nu N^{\nu}&=&0
\eea
and making use of \eq{number} and \eq{cond} we find that
\be{2.11}
\partial_\nu U^{\nu}=
-\frac{U^\nu}{n}\partial_\nu n\ .
\ee
Using \eq{stress}, \eq{fluidcons}, \eq{2.11} and the second law of 
thermodynamics,
we can write a formula for the rate of entropy production using the 
total entropy current four-vector 
$S^\mu\equiv nsU^\mu -U_\nu T_{\mbox{{\tiny TP}}}^{\mu\nu}/T$, where
$s$ is entropy per particle and $T$ the temperature:
\be{2.14}
\partial_\mu S^\mu =-\frac{1}{T}(\partial_\nu U_\mu)
T_{\mbox{{\tiny TP}}}^{\mu\nu}+\frac{1}{T^2}(\partial_\nu T)
U_\mu T_{\mbox{{\tiny TP}}}^{\mu\nu}.
\ee
Now the task is to construct $T_{\mbox{{\tiny TP}}}^{\mu\nu}$ in the way that
the rate of entropy production per unit volume given by \eq{2.14} is positive
for all possible fluid configurations. As explained in \cite{10}, 
we may take the dissipative term to be linearly dependent on the space-time
derivatives of the four-velocity, energy and number densities. Also, the 
effects considered here are of first order and therefore the adiabatic
equations of motion can be used to find out the form of 
$T_{\mbox{{\tiny TP}}}^{\mu\nu}$. 

It is easiest to continue in the locally moving Lorentz-frame, where
$U^\mu =(1,0,0,0)$. As explained, $T_{\mbox{{\tiny TP}}}^{\mu\nu}$ can now
be constructed as a linear combination of the derivatives $\partial U^\mu /
\partial x^\nu$, $\partial U^\mu /\partial t$, $\partial T /\partial x^\nu$
and $\partial T/\partial t$.
Of these, $\partial U^0 /\partial x^\nu$ vanishes in this frame and 
$\partial T/\partial t$ can be expressed in terms of $\nabla\cdot \vec{U}$ 
because of adiabacity. Then the most general structure allowed by rotational
and space-inversion is
\bea{2.17}
T_{\mbox{{\tiny TP}}}^{ij}&=&-\eta \left(\frac{\partial U^i}{\partial x_j}+
\frac{\partial U^j}{\partial x_i}-\frac{2}{3}\nabla\cdot\vec{U}\delta^{ij}
\right)
-\zeta\nabla\cdot\vec{U}\delta^{ij} \ , \cr
\null
T_{\mbox{{\tiny TP}}}^{i0}&=&-\kappa\frac{\partial T}{\partial x_i}-\xi
\frac{U^i}{\partial t} \ , \cr
\null
T_{\mbox{{\tiny TP}}}^{00}&=&0.
\eea
By comparing \eq{2.17} with the usual non-relativistic hydrodynamics we can 
now recognize
$\eta$, $\zeta$ and $\kappa$ to be the shear and bulk viscosity and the heat
conduction coefficients, respectively. $\xi$ is a relativistic contribution
with no non-relativistic counterpart. By using \eq{2.17} in \eq{2.14} and
demanding that the rate of entropy production is positive for all fluid
configurations, we get the result $\xi =T\kappa$ together with the fact that all
the transport coefficients are non-negative. 

Again, following \cite{10}, we consider a radiation dominated plasma having 
very short mean free times. The energy-stress tensor to first order in
the free time $\tau$ for such a fluid reads \cite{10}:
\bea{2.32}
T_{\mbox{{\tiny TP}}}^{\mu\nu}=&-&\left( (4b\tau T^3)/(\frac{\partial \rho}
{\partial T})_n\right)\left(\frac{T}{3}\frac{\partial U^\gamma}{
\partial x^\gamma}+U^\gamma\frac{\partial T}{\partial x^\gamma}\right) \cr
&\times&\left(
(\frac{\partial p}{\partial T})_n (g^{\mu\nu}+U^\mu U^\nu)+
(\frac{\partial \rho}{\partial T})_n U^\mu U^\nu\right) \cr
\null
&-&4b\tau T^3\{ \left(2U^\mu U^\nu U^\gamma + \frac{1}{3}g^{\mu\nu}U^\gamma +
\frac{1}{3}\delta^{\mu\gamma}U^{\nu}+\frac{1}{3}\delta^{\nu\gamma}U^\mu
\right)\frac{\partial T}{\partial x^\gamma} \cr 
\null
&+&\frac{T}{15}\left( 6\frac{\partial (
U^\mu U^\nu U^\gamma)}{\partial x^\gamma} 
+g^{\mu\nu}\frac{\partial U^\gamma}{
\partial x^\gamma}+\frac{\partial U^\mu}{\partial x_\nu}+\frac{\partial U^\nu}
{\partial x_\mu}\right)\} + O(\tau^2),
\eea
where $b$ is a constant. Since $T_{\mbox{{\tiny TP}}}^{\mu\nu}$ is 
a tensor, it is sufficient to study it in the locally comoving Lorentz frame.
Using once again the properties of this frame, $U^\mu =(1,0,0,0)$ and $\partial
U^0/\partial x^\mu =0$, we can somewhat simplify the energy-stress tensor, 
\eq{2.32}:
\bea{2.39}
T_{\mbox{{\tiny TP}}}^{ij}&=&-4bT^4\tau \left(\frac{1}{3}-(\frac{\partial p}
{\partial
\rho})_n\right)^2\nabla\cdot\vec{U}-\frac{4}{15}bT^4\tau \left(
\frac{\partial U^i}
{\partial x_j}+\frac{\partial U^j}{\partial x_i}-\frac{2}{3}\delta^{ij}\nabla
\cdot\vec{U}\right) +O(\tau^2), \cr
\null
T_{\mbox{{\tiny TP}}}^{i0}&=&\frac{4}{3}bT^3\tau \left(\frac{\partial T}
{\partial x_i}+
T\frac{\partial U^i}{\partial t}\right) +O(\tau^2), \cr
\null
T_{\mbox{{\tiny TP}}}^{00}&=&O(\tau^2).
\eea
From \eq{2.39} it is easy to extract the values for the transport coefficients:
\bea{2.43}
\zeta&=&4bT^4\tau \left(\frac{1}{3}-(\frac{\partial p}{\partial \rho})_n
\right)^2, \cr
\eta&=&(4/15)bT^4\tau \ \ \mbox{and} \cr
\kappa&=&(4/3)bT^3\tau .
\eea
As can be seen from \eq{2.43}, the transport coefficients are indeed 
proportional to the mean free time (or equivalently the mean free path) of 
particles 
in the plasma, in which interactions are sufficiently strong so that 
particles have a high scattering frequency.
For various species of particles the coefficients are
obviously different because their cross sections vary and thus they have 
different free paths in the plasma. In Sections 4 and 5 we calculate  
heat conductivity and
shear viscosity considering all the particles and their scattering reactions
in the hot primordial plasma. We note that the values obtained in Sections
4 and 5 are self-consistent with the assumption of the high scattering
frequency, i.e. the mean free times estimated from the values of Table 5
are very short. Let us finally note that 
because the plasma considered here is highly relativistic, and therefore
$p\simeq \frac{1}{3}\rho$, the bulk viscosity $\zeta$ is approximately zero.
%%%%%%%%%%%%%%%%%%%%%%%%%%%%%%%%%%%%%%%%%%%%%%%%%%%%%%%%%%%%%%%%%%%%%%%%%%%%%%
\subsection{The Boltzmann equation}
%%%%%%%%%%%%%%%%%%%%%%%%%%%%%%%%%%%%%%%%%%%%%%%%%%%%%%%%%%%%%%%%%%%%%%%%%%%%%%
To actually calculate the transport coefficients, we need some tool to
treat all the scattering reactions in the plasma. This tool is
the Boltzmann equation, which in the expanding Robertson-Walker universe
reads:
\be{boltzmann}
\frac{\partial f}{\partial t}p^{0}+
p^i \left(\frac{\partial f}{\partial x^i}-2\frac{\dot{R}}{R}p^0
\frac{\partial f}{\partial p^i}-e\frac{\partial f}{\partial p^0}E^i\right)
-\frac{\partial f}{\partial p^0}p^2
R\dot{R}-e\frac{\partial f}{\partial p^i}p^0E^i=C(p,t)p^0~,
\ee
where $C(p,t)$ is the collision integral.
Here we have assumed a Robertson-Walker metric with a scale factor $R$
for the background, and we have defined $F_i^0=-E_i$ and 
$F^{ij}=\epsilon^{ijk}B_k$. The Robertson-Walker metric is valid here
if the mean free path of the particles is sufficiently small i.e. no large
flows are present in the universe. The resulting values in Sections 4 and 5
suggest that the usage of this metric is a self-consistent assumption.
We also prefer to use co-moving coordinates and define
\be{co-mo}
\tilde{f}(p,t)=\int\delta
(p_{0}-(\vec{p}^2R^2+m^2)^{1/2})f(\vec{p},p_{0},t) dp_{0}~.
\ee
Inserting this into \eq{boltzmann},
making use of the local momentum defined as $\tilde\vec p=R\vec p$
and assuming adiabatic expansion in which $E^i\gg (\dot{R}/R) \ |\vec{p}|
\equiv H|\vec{p}|$ we can use the Lorentz-frame and write \eq{boltzmann} 
in the form
(from now on we drop the tildes and we also assume that the electric field
is sufficiently small, $E^i\ll T^2$)
\be{perusyht}
\frac{\partial f}{\partial t}+\frac{p^i}{p_0}\frac{\partial f}{\partial x^i}-
{e}\frac{\partial f}
{\partial p^i}E^i=C(p,t)~.
\ee
To get a sufficiently good estimate, it is only necessary to consider  
2$\rightarrow$2 reactions, and thus we may use as the collision integral
the following expression:
\be{collint}
C(p,t)=\frac{1}{p_0}\int dP_{b}dP_{c}dP_{d}(2\pi )^4
\delta^4 (p+p_b-p_c-p_d)|M(Ab\to cd)|^2{\cal F},
\ee
where $p$ is the four-momentum of the charged test particle  $A$, and
we have defined
$dP_{i}\equiv {d^3 \vec p_i}/({(2\pi )^3}{2E(p_i)})$. The factor
${\cal F}$ contains the distribution functions of the initial and final
states. The Pauli blocking factors must be included for the fermions and the
enhancement factors for the bosons. Below are two examples of the
${\cal F}$-factor. The
first is for the case when all the particles in the reaction
are fermions, and in the second
case particles $b$ and $c$ are bosons, such as is the case for
Compton scattering:
\be{fermi}
{\cal F}=\cases{{
[1-f(p)][1-f_b(p_b)]f_c(p_c)f_d(p_d)-[1-f_c(p_c)][1-f_d(p_d)]
f(p)f_b(p_b)~,}\cr
{[1-f(p)][1+f_b(p_b)]f_c(p_c)f_d(p_d)-
[1-f_d(p_d)][1+f_c(p_c)]f(p)f_b(p_b)~.}\cr}
\ee
The time dependence is not shown explicitly above.
Let us now assume that there exists a constant electric field in the
early universe (Of course, we do not claim such a field actually exists
but rather use it as a probe of the plasma properties. Furthermore, we take
it to be constant in the sense that its coherence length is 
larger than the mean free path of charged particles) and treat its effect  
as a perturbation on the distribution of the test particle. We write
$f=f_{0}+\delta f$, where $f_{0}$ is
the equilibrium
distribution and $\delta f$ the small perturbation. Inserting this into
the collision integral \eq{collint} results in 
\bea{pertcoll}
C(p,t)&=&-\frac{1}{p_0}\int dP_{b}dP_{c}dP_{d}(2\pi )^4\delta^4 
(p + p_b -p_c -p_d)\cr
&\times&|M|^2\big([f_{0c}f_{0d}[1-f_{0b}]+f_{0b}[1-f_{0c}][1-f_{0d}]\big)
\delta f~
\equiv\hat C(p)\delta f(p,t).
\eea
\eq{pertcoll} assumes that all the particles in the reaction are fermions;
generalization to other cases is straightforward.

We have assumed that all the particles b, c and d have an equilibrium 
distribution. Therefore, if also 
the test particle has an equilibrium distribution,
the collision integral vanishes and only the
perturbation term in \eq{pertcoll} survives. 
Note that the equilibrium distribution depends only on the momentum. 

The Boltzmann equation can now be
linearized in order to calculate the electrical conductivity.
Assuming that $\delta f$ depends only on $p$, the linearized Boltzmann 
equation reads
\be{linbo}
v^i\frac{\partial f_0}{\partial x^i}-e\frac{\partial f_0}{\partial p^i}E^i=
\hat C(p)\delta f(p),
\ee
where $v^i$ is the velocity of the probe particle a.
Using \eq{linbo} we can now begin to calculate the transport coefficients
in the early universe. 
%%%%%%%%%%%%%%%%%%%%%%%%%%%%%%%%%%%%%%%%%%%%%%%%%%%%%%%%%%%%%%%%%%%%%%%%%%%%%%
\section{Eletrical conductivity}
%%%%%%%%%%%%%%%%%%%%%%%%%%%%%%%%%%%%%%%%%%%%%%%%%%%%%%%%%%%%%%%%%%%%%%%%%%%%%%
There has been many
attempts to estimate $\sigma$. In \cite{2} this was done in the relaxation
time approximation and in \cite{3} and \cite{4} a Coulomb correction was 
applied. Recently a numerical work was carried out in \cite{1} which showed 
that
the electrical conductivity in the early universe is mainly due to the
leptonic contribution and its value for $T\leq 100$ MeV $\sigma\simeq
0.76T$ while at $T\simeq M_W$ it was found that $\sigma\simeq 6.7T$.

However, in \cite{1} the contribution from lepton-quark scatterings was 
neglected, which
was correctly pointed out in \cite{5}. We have now included these
scatterings and thus acquired a better estimate for the electrical 
conductivity.  The additional reactions now included are of the form 
$l^{\pm}q\rightarrow \l^{\pm}q$ and $l^{\pm}\bar{q}\rightarrow 
\l^{\pm}\bar{q}$, and these affect only the leptonic contribution because
the quarks interact via the strong interaction and thus the electromagnetic
scattering processes mentioned above can still be neglected when considering 
the electrical conductivity created by the quarks.

The $t$- and $u$-channels give singular contributions to the collision
integral and that is why one needs regularization. The full finite
temperature calculations with higher order Feynman graphs and resummation
would provide the regularization correctly but the expressions for the matrix
elements become quite long. Another technical complication is the correct
handling of the real and imaginary parts. A suitably accurate calculation can
be, however, performed by using thermal masses as regulators in the $t$-
and $u$-channel propagators. This approach can be expected to give roughly 
the same results as the one using the full thermal propagator. 

The thermal masses used in the $t$- and $u$-channels for the
lepton, quark, photon and gluon propagators, respectively, are the following
\cite{6}:
\bea{masses}
m_l^2&=&\frac{e^2}{8}T^2\simeq 0.0115 T^2,\cr
m_q^2&=&\frac{g_{s}^{2}}{6}T^2\simeq 0.251 T^2 ,\cr
m_\gamma^2&=&\frac{e^2}{3}(\Sigma_{l}Q_{l}^2+3\Sigma_{q}Q_{q}^{2})T^2\simeq
0.0306(\Sigma_{l}Q_{l}^2+3\Sigma_{q}Q_{q}^{2})T^2,\cr
m_g^2&=&g_{s}^{2}(3+\frac{N_{f}}{3}) T^2 \simeq 
1.508(3+\frac{N_{f}}{3})T^2.
\eea
where $N_{f}$ is the number of quark families present, the sums are over all 
particles with
$m\le T$, and we adopt the values $g_{s}^{2}=4\pi\alpha_s\simeq 1.508$
and $e^2=4\pi\alpha_{e}\simeq 0.0917$, and $\qcd=200 \MeV$.

The matrix elements for all the reactions used in the collision integral
can be found in Tables 1, 2 and 3. All the elements are summed over final
spins and averaged over initial spins; the QCD-processes are also
summed over final color and averaged over initial color. Because we can treat
every spin and color state as a different particle in the initial state we
must, however, remove the averaging in the collision integral by multiplying
the matrix elements with the number of initial spin and color states.

The proper introduction of the 
thermal regulators in the propagators of the $t$- and $u$-channels should 
also be given special attention. 
For example, the matrix element for the reaction $e^-e^+\rightarrow e^-e^+$ 
is the sum $(u^2+s^2)/t^2+(u^2+t^2)/s^2+2u^2/(st)$, 
which in \cite{1} was first simplified into the form 
$(s^4+t^4+u^2(s+t)^2)/(s^2t^2)$ and only after that the regulators were 
inserted into the
$t$- and $u$-channels. But one should be extremely careful when applying 
regularization because
the adding of the regulators in different stages of the calculation changes 
the actual value
of the matrix element, and therefore the value of the conductivity. This is 
so because the regulators are added only to the $t$- and $u$-propagators, 
not consistently everywhere in the matrix element.  
The proper way
to introduce the screening terms is to apply the regulators to every
term given by the Feynman graphs separately and before joining the 
terms together. The regulating should be done this way because the different
terms in the matrix elements have singularities of different order and hence
the terms should be regulated and integrated separately. 

We consider the temperature interval from about $1$ MeV to about $10$ GeV and
make the simple approximation that particles appear in the thermal bath when
temperature is greater than their mass. The collision integral is performed
numerically by evaluating the integral by a simple Monte Carlo
simulation. 

The results, however, are not very different from the previous ones 
\cite{1}; the quark 
contribution is still negligible compared with the leptonic electrical
conductivity and the 
actual value of the conductivity is still $\simeq 10T$ in natural units.
The effect of neutrino scatterings to the contibution of either QED or QCD 
electrical conductivity is negligible. The results are given in Fig. 1 and
in Table 5. 

The main difference between the value for the electrical conductivity here and
in \cite{5} is due to the difference in the definition of the electrical
conductivity. In \cite{5} an electric field extending itself over the whole
universe is used to probe the electrical conductivity of the early universe.
Because in \cite{5} a homogenous field is assumed to cover the entire space,
the leptons (quarks)
and antileptons (antiquarks) are accelerated in the opposite directions. 
In this scenario, it is justified to neglect in the first order all other
scattering reactions than the lepton-antilepton (quark-antiquark) annihilation.
But the resulting electrical conductivity is a global one - one that is
the same in the whole universe and does not take into consideration the
small patch-like structure of the primordial magnetic field.

The primordial magnetic field $\vec{B}$, if it exists, is most likely
random: both its magnitude and direction vary.
Therefore in the early universe it is more sensible to 
talk about the local electrical conductivity, $\sigma (\vec{x})$. 
The magneto hydrodynamic equation reads then
\be{hydris}
\frac{\partial\vec{B}}{\partial t}=\nabla\times (\vec{v}\times\vec{B})-
\nabla\times \left( \frac{\nabla\times\vec{B}}{\sigma (\vec{x})}\right) .
\ee
From \eq{hydris} we see that to blindly treat the electrical conductivity 
strictly as a constant would disregard some aspects of the dissipation of 
the magnetic field. However,
to a good approximation we can take the electrical conductivity to be a 
constant inside the patches of the magnetic field characterized by correlation 
length $L$. 

The magnetic field is coherent inside the correlation length $L$.
The different patches 
containing magnetic field are uncorrelated and thus a convenient way to
define the local conductivity is the measure of the dissipation of the 
magnetic fields in the coherent patches. We measure the local 
electrical conductivity as experienced by a charged particle shot out of 
one of the 
magnetic patches to the next, uncorrelated one. The particles are 
ultrarelativistic because of the 
high temperature and therefore the different bulk velocities between the
magnetic patches can be neglected. 

The condition when this scenario is valid can be easily checked. The mean free
path of a particle in the plasma is $l_{free}=(\sigma_{tp} n_p)^{-1}$, where 
$\sigma_{tp}$ is the transport cross section of the particle species $p$, and
$n_p$ 
is its number density. The condition for validity is now that
the mean free path is longer than the coherence length of the magnetic
field, $L<l_{free}$. Typical coherence length could be $L\simeq 1/T$ whereas
$l_{free}\simeq 1/(\alpha^2 T)$. In this case particles will typically be 
shot out of a given patch before they have time to interact. When they 
finally do,
this will happen in a different patch where the other particles have a 
bulk motion of their own.

However, the coherence length of the primordial magnetic field is not 
known exactly.
It might very well be of the order of the natural microphysical scale, the
interparticle distance, but there
could be processes and phenomena that would cause it to be much 
larger. For example, one such process studied recently is the inverse 
cascade \cite{25}, which transfers magnetic energy from small length 
scales to larger length scales.

Here we shall assume that the mean free path of a particle is longer than
the correlation length of the primordial magnetic field.
To calculate the local conductivity, we assume one test particle
from each species of particles being shot out of one patch of magnetic field 
to another, totally uncorrelated one. In the new patch, we assume a patch-wide
electric field to give us a tool to measure the conductivity. Of course, this
is just a test field, we do not claim such fields to exist in the early 
universe. Inside the patch, we also assume isotropy and therefore we can
write \eq{linbo} to the form
\be{ellinbo}
-e\frac{\partial f_0}{\partial p^i}E^i=\hat C(p)\delta f(p).
\ee
From \eq{ellinbo} one easily obtains 
\be{delta}
\delta f(p)=-\frac{e}{\hat C(p)}\frac{\partial f_0(p)}{\partial p^i}E^i~.
\ee
The induced current density is given by
\be{jii}
\vec j=e\int\frac{\delta f(p)\vec{p}}{p}d^3\vec{p}~,
\ee
and conductivity $\sigma_A$, associated with a given particle species $A$, 
is defined by
\be{condi}
\vec j_A={\sigma_A}\vec E~.
\ee
Thus the contribution of a single species $A$ to conductivity is
$\sigma_A\sim 1/\sum\vert M(AX\to Y)\vert^2$, where the sum is over
all the processes which scatter the test particle $A$. For the purpose
of conductivity, we may view the mixture of different particle species of
the early universe a multicomponent fluid. The flow of each
component contributes to the total current and adds up to the total 
conductivity, which reads
\be{condtot}
\sigma_{\rm tot}=\sum_A\sigma_A~,
\ee
where the sum is over all the relativistic charged species present in
the thermal bath. Note that the
total conductivity is dominated by the species that has the weakest
interaction. This is because the weaker the interaction, the longer time
the current flow  is maintained.

We consider the temperature interval $1\MeV \lsim T\lsim 10\GeV$, and
make the simple, crude assumption that particles appear in the thermal
bath only when temperature is greater than their mass. Thus below
$100$ MeV, for example, the only charged particles present are the
electrons and positrons.
When $T\gsim \qcd$, also the quarks should be counted in. Their
main interactions are strong, so that their electromagnetic interactions
may be neglected. The list of the relevant reactions involving QED and
QCD charged particles can be found in Tables 1 and 2.

\begin{figure}
\leavevmode
\centering
\vspace*{80mm}
\includegraphics{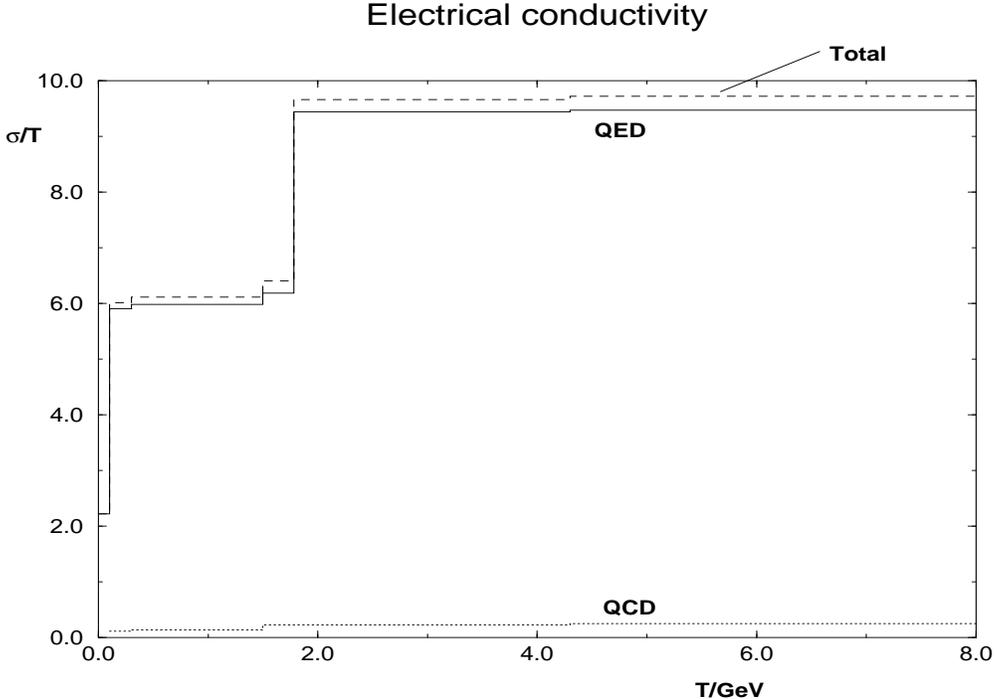}
\caption{$\sigma /T$ as a function of temperature.}
\label{kuva1}
\end{figure}

%%%%%%%%%%%%%%%%%%%%%%%%%%%%%%%%%%%%%%%%%%%%%%%%%%%%%%%%%%%%%%%%%%%%%%%%%%%%%%
\section{Thermal conductivity}
%%%%%%%%%%%%%%%%%%%%%%%%%%%%%%%%%%%%%%%%%%%%%%%%%%%%%%%%%%%%%%%%%%%%%%%%%%%%%%
To find out the thermal conductivity, we now abandon the isotropy requirement
and the requirement of locality used in the previous Section in computing
the electrical conductivity.
Let us for a while let the equilibrium distribution
$f_0$ depend also on the chemical potential $\mu$: $f_0=1/(1\pm e^{(p-\mu )/T}
)$. The conditions for thermal equilibrium require the constancy of temperature
and of the sum $\mu + V$, where $V$ is the energy of the 
particles in an external field, throughout the media considered.  Here we 
take $V=e\phi$, where
$\phi$ is the electric field potential. In a plasma with a non-uniform
temperature distribution, the electric field $\vec{E}$ is not zero even if
the current is zero. In general, when both the current density $\vec{j}$
and the temperature gradient $\nabla T$ are not zero, the relation between
these quantities and the electric field can be written as
\be{seliseli}
\vec{j}=\sigma\vec{E}-\alpha\sigma\nabla T+\frac{\sigma\nabla\mu}{e},
\ee
where $\sigma$ is the electrical conductivity and $\alpha$ the thermoelectric
coefficient.
We can now write the Boltzmann transport equation 
in the form
\be{boltz2}
-e\vec{E}\cdot \frac{\partial f_0}{\partial \vec{p}}+\frac{\vec{p}}{p}\cdot
\frac{\partial f_0}{\partial \vec{r}}=\hat C(p,T)\ \delta f.
\ee
Assuming 
that the temperature $T$ and the chemical potential $\mu$ are functions of 
the spatial coordinates 
it is easy to solve $\delta f$, the small perturbation 
to the equilibrium distribution $f$ from \eq{boltz2}:
\bea{deltaf}
\delta f&=&\frac{\vec{p}}{p\hat C(p,T)}\cdot\{ -e\vec{E}\frac{\partial f_0}
{\partial p}+
\frac{\partial f_0}{\partial \mu}\nabla\mu +\frac{\partial f_0}
{\partial T}\nabla T\} \cr
\null
&=&\frac{\pm e^{(p-\mu )/T}}{(1\pm e^{(p-\mu )/T})^2}\frac{\vec{p}}
{pT\hat C(p,T)}\cdot\{ e\vec{E}+\nabla\mu +(p-\mu )\frac{\nabla T}{T}\}
\eea
In order to determine the thermoelectric coefficient $\alpha$, we must 
now assume for a while that
the condition $e\vec{E}+\nabla\mu =\vec{0}$ holds. From the definition of the
electric current \eq{jii} and from \eq{seliseli} one gets
\be{onsa}
\vec{j}=-\alpha\sigma\nabla T=-e\int\frac{\vec{p}\delta f}{p}d^3\vec{p}.
\ee
Remembering the earlier definition of $\sigma$, \eq{jii} and \eq{condi}, it 
is now easy to solve the thermoelectric coefficient:
\be{thermoe}
\alpha =\frac{1}{eT}(\mu -G),
\ee
where 
\be{G}
G\equiv \int\frac{p^3e^{(p-\mu )/T}dp}{(1\pm e^{(p-\mu )/T})^2\hat C(p)} \ 
\times\ \left(  
\int\frac{p^2e^{(p-\mu )/T}dp}{(1\pm e^{(p-\mu )/T})^2\hat C(p)}\right)^{-1}.
\ee
Now let us assume that $\vec{j}=\vec{0}$, in which case we obtain from 
\eq{seliseli} that $\alpha\nabla T=\vec{E}+\nabla\mu /e$ and that
\be{delta3}
\delta f=\frac{\pm e^{(p-\mu )/T}}{(1\pm e^{(p-\mu )/T})^2}\frac{p-G}{T^2C(p)}
\frac{\vec{p}\cdot\nabla T}{p}.
\ee
From the energy flux relation 
\be{q}
\vec{q}=\int\vec{p}\delta f d^3\vec{p}\equiv -\kappa\nabla T,
\ee
and from noting that in the early universe we can put the chemical potential
$\mu$ to zero, we get an expression for the thermal conductivity coefficient
$\kappa$:
\be{kappa}
\kappa = \frac{\pm 4\pi}{3T^2}\int_{0}^{\infty}\frac{p^3}{\hat C(p)}
\frac{e^{p/T}}{(1\pm e^{p/T})^2}(G-p)dp.
\ee
The plus-minus sign refers to fermions and bosons, respectively. The heat
conductivity coefficient is, however, always positive because the
${\cal F}$ (\eq{fermi}) factor also changes its sign in the collision 
integral when considering fermions or bosons, respectively.

As explained in Section 2, thermal conductivity is related to 
the mean free time, or equivalently, to the mean free path. 
The values for the heat conductivity obtained in this Section are 
self-consistent with the assumption of short mean free times in Section 2
as can be seen by comparing the values presented in Table 5 and \eq{2.43}.

From \eq{kappa} it can be easily seen that $\kappa$ is proportional
to the inverse of the square of the coupling constant of the scattering 
reaction in question.
Thus the smaller the coupling constant is the bigger the value for $\kappa$
will be i.e. $\kappa$ indeed is related to the mean free path.

The subtle interplay between the increasing effect on thermal conductivity
by the temperature
dependent regulators and the decreasing effect due to the change in
the amount of particles (potential scatterers) present should also be noted. 
As can be seen in
Fig. 2 and Table 5, the case for the leptonic thermal conductivity, 
$\kappa_{l}$, is clear. The first effect dominates 
in the lepton case, increasing $\kappa_{l}/T^2$ all the way from the case 
where only electrons are present to the case where also the b-quark is present
in the heat bath. The photon case is also straightforward. Because there 
are no 
temperature dependent regulators the only effect comes from the fact that
more and more scatterers are present the higher the temperature is. Thus we
see in Fig. 2 and Table 5 that the photonic thermal conductivity divided by 
the temperature
squared, $\kappa_{\gamma}/T^{2}$, is decreasing steadily as the temperature 
rises.

\begin{figure}
\leavevmode
\centering
\vspace*{80mm}
\includegraphics{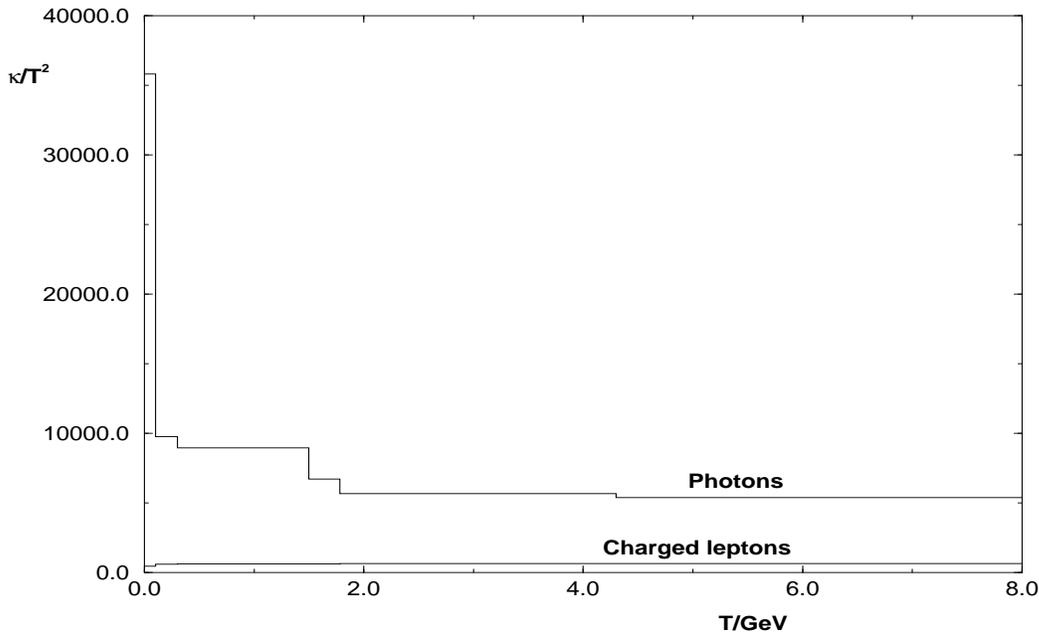}
\caption{$\kappa /T^2$ as a function of temperature for leptons and photons.}
\label{kuva2}
\end{figure}
The case is similar with gluons and quarks as can be seen 
in Fig. 3 and Table 5. 
The different behaviour of the quark thermal conductivity $\kappa_{q}$ 
and the gluonic thermal conductivity $\kappa_{g}$ can be 
understood when one looks at
the different processes that are involved in the scatterings. All QCD matrix 
elements are listed in Table 2. The quarks get the biggest 
contribution to 
their collision integral from the potentially singular matrix element which 
has been regulated by thermal masses while
the gluons are not that sensitive to thermal masses.

The thermoelectric coefficient $\alpha$ in \eq{thermoe} for leptons is found 
to be $\pm 39.1$ (depending on the sign of the lepton electric charge) and
for quarks $47.4/e_q$, where $e_q$ is the electric charge of the quark.

\begin{figure}
\leavevmode
\centering
\vspace*{80mm}
\includegraphics{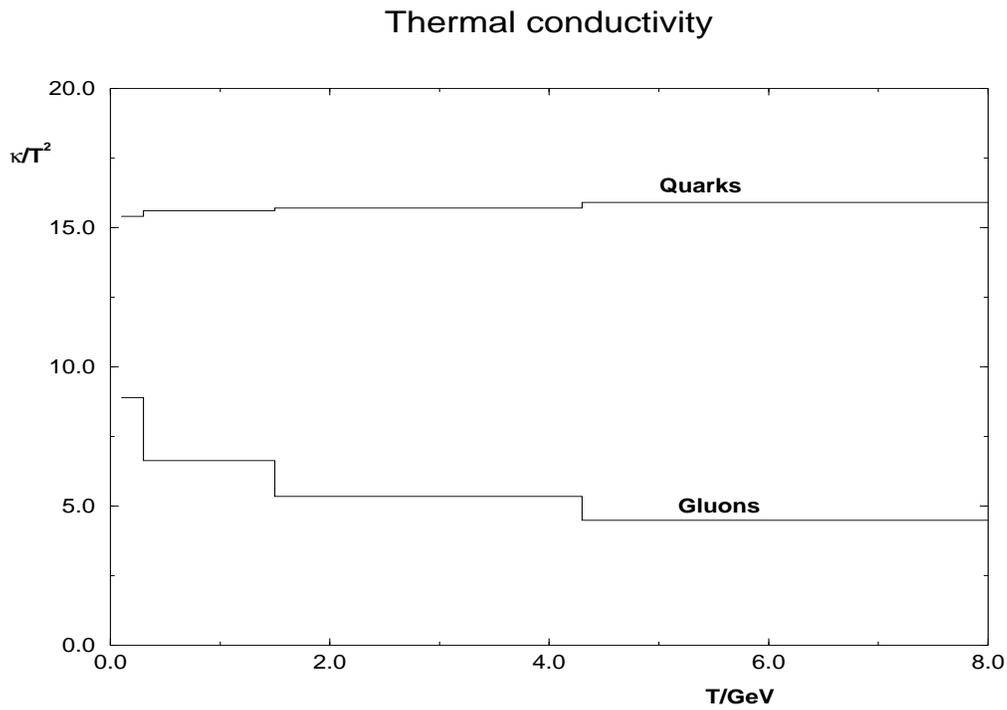}
\caption{$\kappa /T^2$ as a function of temperature for quarks and gluons.}
\label{kuva3}
\end{figure}
The neutrinos do not have any singularities in their matrix elements and thus
as the temperature grows and more and more
scatterers appear in the heat bath the neutrino thermal conductivity 
divided by the temperature squared, $\kappa_\nu /T^2$, decreases steadily. 
$\kappa_{\nu}/T^2$ is shown in Fig. 4.

Earlier attempts to estimate the heat conductivity can be found, for example,
in \cite{11,8,9}. However, in \cite{11} no actual matrix
elements were calculated but
heat conductivity was estimated to be proportional to the mean free time
of a particle species studied.
In \cite{8} and \cite{9} a more thorough calculation was performed to the
leading logarithmic order. However, due to the approximations used in the
formalisms of \cite{8} and \cite{9} we believe that values in the present
paper are more accurate.

\begin{figure}
\leavevmode
\centering
\vspace*{80mm}
\includegraphics{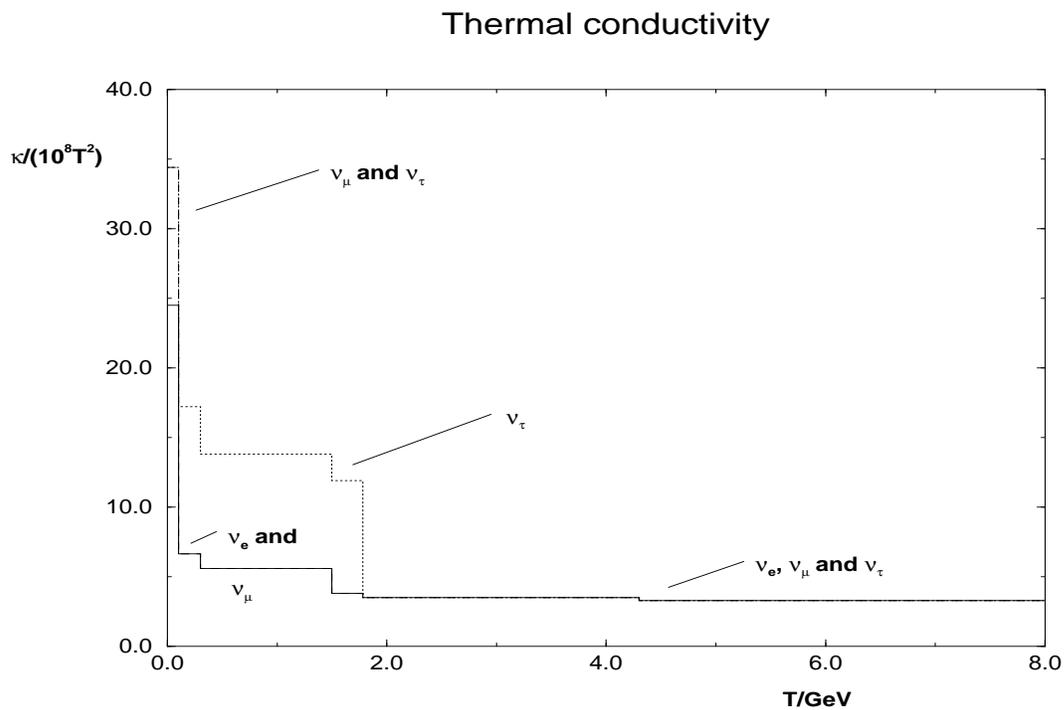}
\caption{$\kappa /T^2$ as a function of temperature for neutrinos.}
\label{kuva4}
\end{figure}

{\begin{table*}
\centering
\caption[t1]{Matrix elements $|M|^2/(32\pi^2\alpha_{em})$ for the QED 
processes used in the 
collision integral. All matrix elements are summed over final 
spins and averaged over initial spins. $\alpha_{em}$ is the electromagnetic
coupling constant, $i\neq j$, and $e_q$ is the charge of the quark in 
question.}
\vskip10pt
\begin{tabular}{|c|c|}
\hline
$\mbox{$l$}^{-}\mbox{$l$}^{+}\rightarrow\gamma\gamma$     &  
$ut(\frac{1}{u^2}+\frac{1}{t^2})$ \\
\hline
$\mbox{$l$}^{-}\gamma\rightarrow\mbox{$l$}^{-}\gamma$ & 
$-us(\frac{1}{u^2}+\frac{1}{s^2})$ \\
\hline
$\mbox{$l$}^{-}\mbox{$l$}^{+}\rightarrow\mbox{$l$}^{-}\mbox{$l$}^{+}$ &
$\frac{u^2+s^2}{t^2}+\frac{u^2+t^2}{s^2}+\frac{2 u^2}{st}$ \\
\hline
$\mbox{$l$}^{-}\mbox{$l$}^{-}\rightarrow\mbox{$l$}^{-}\mbox{$l$}^{-}$ &
$\frac{u^2+s^2}{t^2}+\frac{s^2+t^2}{u^2}+\frac{2 s^2}{ut}$ \\
\hline
$\mbox{$l_i$}^{-}\mbox{$l_j$}^{-}\rightarrow\mbox{$l_i$}^{-}\mbox{$l_j$}^{-}$ &
$\frac{s^2+u^2}{t^2}$   \\
\hline
$\mbox{$l_i$}^{-}\mbox{$l_i$}^{+}\rightarrow\mbox{$l_j$}^{-}\mbox{$l_j$}^{+}$ &
$\frac{u^2+t^2}{s^2}$ \\
\hline
$\mbox{$l$}^{-}\mbox{$l$}^{+}\rightarrow q\bar q$ &
$3 e_{q}^{2}\ \frac{u^2+t^2}{s^2}$ \\
\hline
$\gamma\gamma\rightarrow\mbox{$l$}^{-}\mbox{$l$}^{+}$     &  
$ut(\frac{1}{u^2}+\frac{1}{t^2})$ \\
\hline
$\gamma\gamma\rightarrow q\bar q$     &  
$3e_{q}^{2}\ ut(\frac{1}{u^2}+\frac{1}{t^2})$ \\ 
\hline
$\gamma\mbox{$l$}^{-}\rightarrow\gamma\mbox{$l$}^{-}$ & 
$-us(\frac{1}{u^2}+\frac{1}{s^2})$ \\
\hline
$\gamma q\rightarrow\gamma q$ & 
$-3e_{q}^{2}\ us(\frac{1}{u^2}+\frac{1}{s^2}) $ \\
\hline
\end{tabular}
\label{t1}
\end{table*}}
%\vskip30pt

{\begin{table*}
\centering
\caption[t2]{Matrix elements $|M|^2/(16\pi^2\alpha_{s})$ for the QCD processes 
\cite{26} used in the collision integral. All matrix elements are summed 
over final spins and colors and averaged over initial spins and colors.
$\alpha_{s}$ is the strong coupling constant, and $i\neq j$.}
\vskip10pt
\begin{tabular}{|c|c|}
\hline
$\mbox{$q$}\bar{\mbox{$q$}}\rightarrow gg$  &
$\frac{8}{3}(\frac{4}{9}(\frac{ut}{t^2}+\frac{ut}{u^2})-2(\frac{
s^2-ut}{s^2})-(\frac{u^2}{us}+\frac{t^2}{st}))$ \\
\hline
$\mbox{$q$}g\rightarrow\mbox{$q$}g$ &
$2(\frac{t^2-us}{t^2})-\frac{4}{9}(\frac{us}{u^2}+\frac{us}{s^2})+
(\frac{u^2}{ut}+\frac{s^2}{st})$ \\
\hline
$\mbox{$q$}\bar{\mbox{$q$}}\rightarrow\mbox{$q$}\bar{\mbox{$q$}}$ &
$\frac{4}{9}(\frac{u^2+s^2}{t^2}+\frac{u^2+t^2}{s^2}-\frac{2u^2}{3st})$ \\
\hline
$\mbox{$q$}\mbox{$q$}\rightarrow\mbox{$q$}\mbox{$q$}$ &
$\frac{4}{9}(\frac{u^2+s^2}{t^2}+\frac{s^2+t^2}{u^2}-\frac{2s^2}{3ut})$ \\
\hline
$\mbox{$q_i$}\mbox{$q_j$}\rightarrow\mbox{$q_i$}\mbox{$q_j$}$ &
$\frac{4}{9}\frac{s^2+u^2}{t^2}$   \\
\hline
$\mbox{$q_i$}\bar{\mbox{$q_i$}}\rightarrow\mbox{$q_j$}\bar{\mbox{$q_j$}}$ &
$\frac{4}{9}\frac{u^2+t^2}{s^2}$ \\
\hline
$gg\rightarrow\mbox{$q$}\bar{\mbox{$q$}}$  &
$\frac{3}{8}(\frac{4}{9}(\frac{ut}{t^2}+\frac{ut}{u^2})-2(\frac{
s^2-ut}{s^2})-(\frac{u^2}{us}+\frac{t^2}{st}))$ \\
\hline
$g\mbox{$q$}\rightarrow g\mbox{$q$}$ &
$2(\frac{t^2-us}{t^2})-\frac{4}{9}(\frac{us}{u^2}+\frac{us}{s^2})+
(\frac{u^2}{ut}+\frac{s^2}{st}))$ \\
\hline
$gg\rightarrow gg$ &
$\frac{9}{8}(\frac{17t^2-8us}{2t^2}+\frac{17u^2-8st}{2u^2}+
\frac{17s^2-8ut}{2s^2}+\frac{15ut-s^2}{2ut}+\frac{15ts-u^2}{2st}+
\frac{15us-t^2}{2su}-\frac{135}{4})$ \\
\hline
\end{tabular}
\label{t2}
\end{table*}}
%\vskip30pt

{\begin{table*}
\centering
\caption[t3]{Matrix elements $|M|^2/G^2$ for the neutrino processes 
used in the collision integral. All matrix elements are summed over final 
spins and averaged over initial spins. 
$G$ is the Fermi coupling constant, and $i\neq j$. For other abbreviations,
see Table 4.}
\vskip10pt
\begin{tabular}{|c|c|}
\hline
$\nu_i\nu_i\rightarrow\nu_i\nu_i$ & 
$24s^2$ \\
\hline
$\nu_i\nu_j\rightarrow\nu_i\nu_j$ &
$8s^2$ \\
\hline
$\nu_i\bar{\nu}_i\rightarrow\nu_i\bar{\nu}_i$ &
$24 u^2$ \\
\hline
$\nu_i\bar{\nu}_j\rightarrow\nu_i\bar{\nu}_j$ &
$8 u^2$ \\
\hline
$\nu_i\bar{\nu}_i\rightarrow\mbox{$l$}_i\bar{\mbox{$l$}}_{i}$ &
$8((g_V+g_A+2)^2u^2+(g_V-g_A)^2t^2)$ \\
\hline
$\nu_i\bar{\nu_i}\rightarrow\mbox{$l$}_j\bar{\mbox{$l$}}_j$ &
$8((g_V+g_A)^2u^2+(g_V-g_A)^2t^2)$ \\
\hline
$\nu_i\mbox{$l$}_i\rightarrow\nu_i\mbox{$l$}_i$ &
$4((g_V+g_A+2)^2s^2+(g_V-g_A)^2u^2)$ \\
\hline
$\nu_i\mbox{$l$}_j\rightarrow\nu_i\mbox{$l$}_j$ &
$4((g_V+g_A)^2s^2+(g_V-g_A)^2u^2)$ \\
\hline
$\nu_i\bar{\mbox{$l$}}_j\rightarrow\nu_i\bar{\mbox{$l$}}_j$ &
$4((g_V+g_A)^2u^2+(g_V-g_A)^2s^2)$ \\
\hline
$\nu_i\bar{\mbox{$l$}}_i\rightarrow\nu_i\bar{\mbox{$l$}}_i$ &
$4((g_V-g_A)^2s^2+(g_V+g_A+2)^2u^2)$ \\
\hline
$\nu_i\bar{\mbox{$l$}}_i\rightarrow\nu_j\bar{\mbox{$l$}}_j$ &
$16u^2$ \\
\hline
$\nu_i\mbox{$l$}_j\rightarrow\nu_j\mbox{$l$}_i$ &
$16s^2$ \\
\hline
$\nu_i\bar{\nu}_i\rightarrow q\bar{q}$ &
$24((g_{V}^{q}+g_{A}^{q})^2u^2+(g_{V}^{q}-g_{A}^{q})^2t^2)$ \\
\hline
$\nu_i q\rightarrow\nu_i q$ &
$12((g_{V}^{q}+g_{A}^{q})^2s^2+(g_{V}^{q}-g_{A}^{q})^2u^2)$ \\
\hline
$\nu_i\bar{q}\rightarrow\nu_i\bar{q}$ &
$12((g_{V}^{q}+g_{A}^{q})^2u^2+(g_{V}^{q}-g_{A}^{q})^2s^2)$ \\
\hline 
$\nu_i q_1\rightarrow\mbox{$l$}_iq_2$ &
$48s^2V_{12}$ \\
\hline
$\nu_i\bar{\mbox{$l$}}_i\rightarrow q_1q_2$ &
$48u^2V_{12}$ \\
\hline
\end{tabular}
\label{t3}
\end{table*}}
%\vskip30pt

{\begin{table*}
\centering
\caption[t4]{Abbreviations for Table 3}
\vskip10pt
\begin{tabular}{|c|c|}
\hline
$g_V$ &
$-\frac{1}{2}+2 \sin^2\Theta_W$ \\
\hline
$g_A$ &
$-\frac{1}{2}$ \\
\hline
$g_{V}^{q}(\mbox{charge}\ e_q=+2/3)$ &
$\frac{1}{2}-\frac{4}{3} \sin^2\Theta_W$ \\
\hline
$g_{A}^{q}(\mbox{charge}\ e_q=+2/3)$ &
$\frac{1}{2}$ \\
\hline
$g_{V}^{q}(\mbox{charge}\ e_q=-1/3)$ &
$-\frac{1}{2}+\frac{2}{3} \sin^2\Theta_W$ \\
\hline
$g_{A}^{q}(\mbox{charge}\ e_q=-1/3)$ &
$-\frac{1}{2}$ \\
\hline
$V_{ud}$ &
$0.975$ \\
\hline
$V_{us}$ &
$0.221$ \\
\hline
$V_{ub}$ &
$0.000$ \\
\hline
$V_{cd}$ &
$0.200$ \\
\hline
$V_{cs}$ &
$0.979$ \\
\hline
$V_{cb}$ &
$0.050$ \\
\hline
\end{tabular}
\label{t4}
\end{table*}}
%\vskip30pt
%%%%%%%%%%%%%%%%%%%%%%%%%%%%%%%%%%%%%%%%%%%%%%%%%%%%%%%%%%%%%%%%%%%%%%%%%%%%%%
\section{Viscosity}
%%%%%%%%%%%%%%%%%%%%%%%%%%%%%%%%%%%%%%%%%%%%%%%%%%%%%%%%%%%%%%%%%%%%%%%%%%%%%%
To find out the shear viscosity in the early universe we, once again, set the
stage by starting from the Boltzmann equation \eq{boltzmann} and the 
definition of the shear viscosity and assuming stationary viscous flow:
\be{visco}
-\eta (\partial_iV_k+\partial_kV_i-\frac{2}{3}\delta_{ik}\nabla\cdot\vec{V})
=\int d^3\vec{p}p_iv_k \delta f\ ,
\ee
where $\eta$ is the shear viscosity, $\vec{v}$ the 
velocity of a particle and $\vec{V}$ the velocity of the flow. As in
Section 4, we do not need the requirement of locality used in Section 3
in the computation of the electrical conductivity. $\delta f$
can now be deduced from \eq{linbo} with $\vec{E}=\vec{0}$: 
\be{deltataas}
\delta f=\frac{\nabla f_0\cdot\vec{v}}{\hat C(p)}.
\ee
Substituting \eq{deltataas} into \eq{visco} we see that 
\be{lisavsico}
\int d^3\vec{p}p_iv_k\frac{\nabla f_0\cdot\vec{v}}{\hat C(p)}\simeq
-\int d^3\vec{p}p_iv_k\vec{v}\cdot\nabla (\vec{p}\cdot\vec{V})\frac{\partial
f_0}{\partial p}\frac{1}{\hat C(p)}
\ee
Let us now assume
a flow $\vec{V}=a y \vec{e}_x$, where $a$ is a constant. 
Using $\vec{V}$, \eq{visco} and \eq{lisavsico} we can
write the following expression for the shear viscosity:
\be{visco1}
\eta =\int d^3\vec{p}v_xv_yp_xp_y\frac{\partial f_0}{\hat C(p)\partial p}.
\ee
Using the fact that $\vec{v}$ and $\vec{p}$ have the same directions and
therefore $p_iv_k=p_kv_i$, and
performing the integration in spherical coordinates, we get 
\be{visco2}
\eta =-\frac{1}{T}\int_{0}^{\pi}d\Theta\int_{0}^{2\pi}d\Phi\int_{0}^{\infty}
p^4\sin^2\Phi\cos^2\Phi\sin^5\Theta\frac{\pm e^{p/T}}{\hat C(p)
(1\pm e^{p/T})^2}.
\ee
The $\pm$ sign stands for fermions and bosons, respectively. The shear
viscosity is, however, always found to be positive.
Just as in the case of the thermal conductivity, we see here that the smaller
the interaction of the particle species in question is, the bigger the 
shear viscosity of that species in the plasma is i.e. the viscosity is 
related to 
the mean free time of the particles, too. Here we see a similar behaviour
as with the thermal conductivity in Section 4: the viscosity depends on the
number of scattering reactions  and on the thermal regulators. 
The scattering reactions are the same as in the previous
sections, presented in Tables 1, 2 and 3. In Fig. 6 we show the results for
the shear viscosity of the leptons and photons, and in Fig. 7 the shear 
viscosity of the 
quarks and gluons are presented. In Fig. 8 the neutrino shear viscosity is
presented; all the results are collected in Table 5. It should be noted 
that the
shear viscosities are the same for particles and antiparticles.
Quarks and gluons
interact strongly and thus their shear viscosity is the smallest while,
respectively,
the neutrinos interact weakly so that their shear viscosity is the 
largest. This can be understood by considering the shear viscosity with the 
following example.
Let us shoot a particle into the hot (primordial) plasma. The weaker its
interactions with the surrounding plasma are, the longer it will travel
without scatterings, and vice versa. This is consistent with how we defined 
the shear viscosity in Section 2: it is indeed related to the mean free 
time (the mean free path) of a particle in the plasma. The values for the
shear viscosity obtained in this Section are self-consistent with the
assumption of short mean free times in Section 2 as can be seen by
comparing the values presented in Table 5 and \eq{2.43}.

Earlier estimates for the shear viscosity in hot (primordial) plasma can be 
found, for example, in \cite{10,7,8,9,24}. 
However, in \cite{10} no matrix elements
were calculated, and it was only noted that the shear viscosity is proportional
to the mean free path of a particle species studied.  \cite{7,8,9,24}
contained much more thorough calculations of the shear viscosity to the
leading logarithmic order. The results, however, differ somewhat from the 
ones presented
in Table 5 and Figs. 5, 6 and 7. We believe that the proper introduction
and numerical integration of the matrix elements in the collision integral 
in the present calculation yields more accurate values for $\eta$ than in the 
earlier publications.

As noted before, in the early universe $p\simeq\frac{1}{3}\rho$, so that the 
bulk viscosity (\eq{2.43}) is approximately zero.

\begin{figure}
\leavevmode
\centering
\vspace*{80mm}
\includegraphics{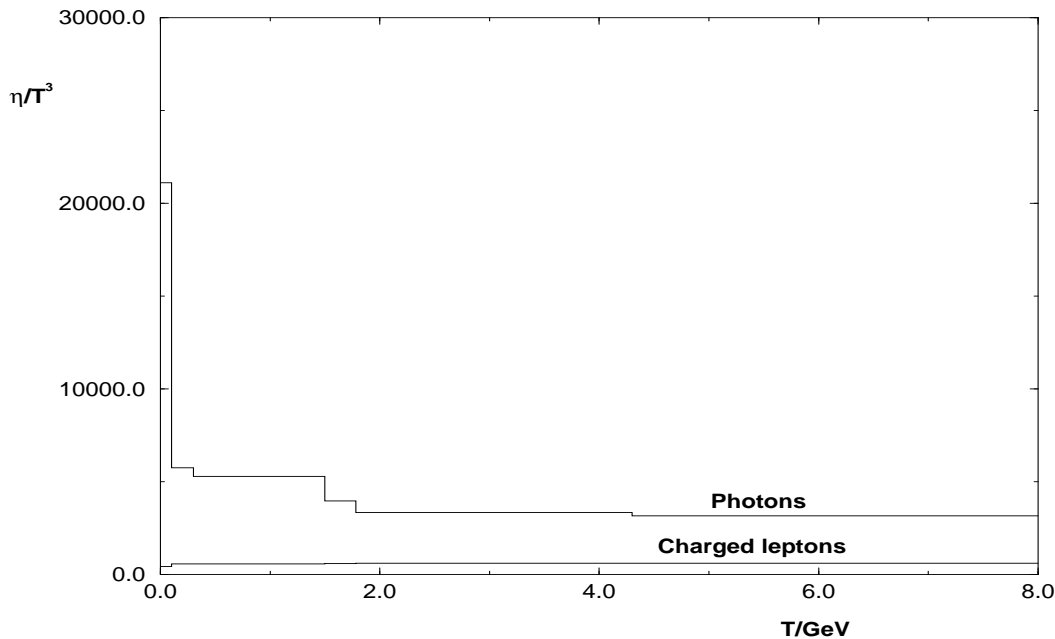}
\caption{$\eta /T^3$ as a function of temperature for charged leptons 
and photons}
\label{kuva5}
\end{figure}

\begin{figure}
\leavevmode
\centering
\vspace*{80mm}
\includegraphics{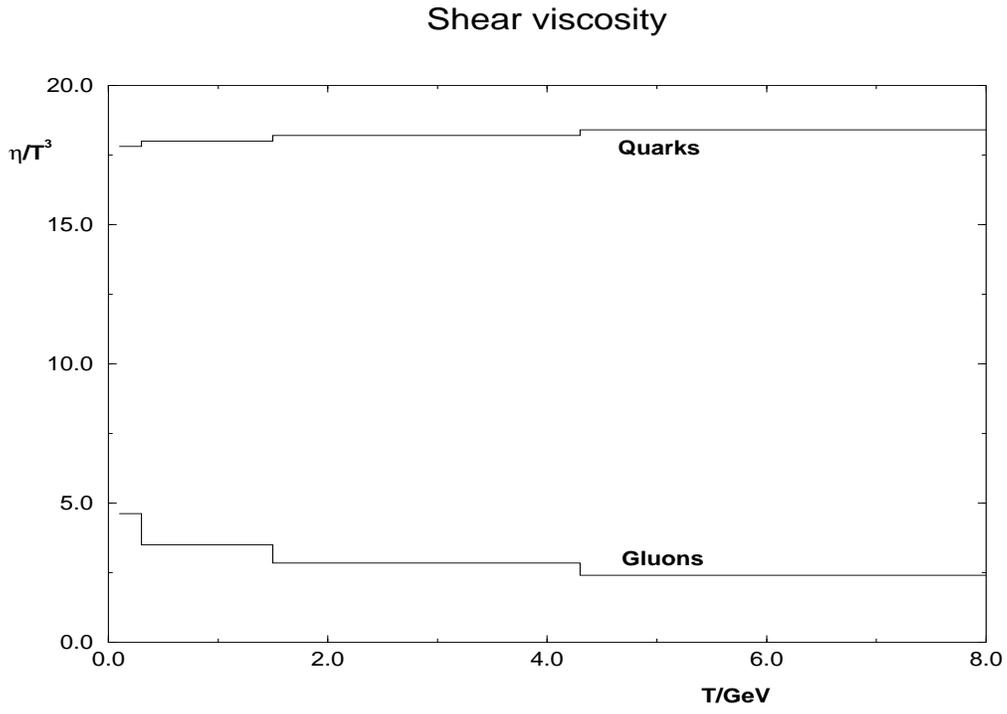}
\caption{$\eta /T^3$ as a function of temperature for quarks and gluons}
\label{kuva6}
\end{figure}

\begin{figure}
\leavevmode
\centering
\vspace*{80mm}
\includegraphics{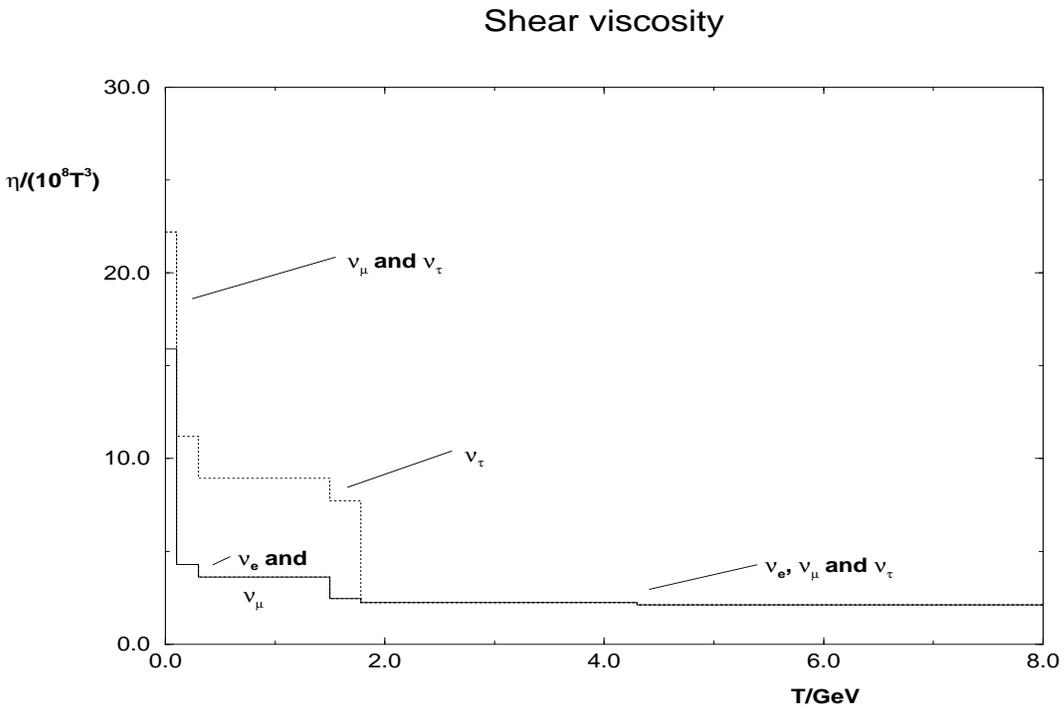}
\caption{$\eta /T^3$ as a function of temperature for neutrinos.}
\label{kuva7}
\end{figure}
%%%%%%%%%%%%%%%%%%%%%%%%%%%%%%%%%%%%%%%%%%%%%%%%%%%%%%%%%%%%%%%%%%%%%%%%%%%%%%
\section{Summary}
%%%%%%%%%%%%%%%%%%%%%%%%%%%%%%%%%%%%%%%%%%%%%%%%%%%%%%%%%%%%%%%%%%%%%%%%%%%%%%
We find that the electrical conductivity is dominated by the leptonic current
and that the quark and neutrino contribution to the electrical conductivity
can be neglected. The values for the electrical conductivity can
be found in Table 5. The definition of the electrical conductivity
should be carefully considered. We started from the magneto hydrodynamic 
equation
\eq{hydris} and noted that the electrical conductivity defines the dissipation
rate of the magnetic field in the plasma. But because of the short correlation
lengths of the magnetic fields in the early universe, the electrical 
conductivity must be a local quantity. The different patches of the 
magnetic field
are uncorrelated and therefore our way of defining and calculating the 
electrical conductivity seems natural: when the mean free path of the particles
in the hot primordial plasma is longer than the correlation length of the
magnetic field, particles will be shot out of the small magnetic patches to
the nearby uncorrelated ones and we can calculate the electrical 
conductivity by
examining how fast a charged particle loses its initial bulk momentum.
To define a global electric field which would extend itself over the whole
universe would disregard the small random structure of the primordial magnetic 
field. We assume in the paper that the mean free path of the particles
studied in the primordial plasma is always longer than the correlation
length of the primordial magnetic field.

To compute the heat conductivity and shear viscosity we use the same tools
as in the computation of the electrical conductivity, but without the 
requirement of locality.
The creation of the primordial magnetic fields can also depend on the viscosity
and heat transportation of the hot primordial plasma \cite{15,16}. 
The viscous damping and 
heat conducting effects do also affect the first order phase transitions in the
early universe \cite{16}. During the phase transitions, 
instabilities may occur when
the transport of latent heat is dominated by the fluid flow. The instabilities
can be damped by finite viscosity and heat conductivity due to the diffusion
of radiation on small length scales. The galaxy formation can also be 
considered to be dependent on the transport coefficients in the early 
universe \cite{13,14}. A summary of the values for the
heat conductivity and shear viscosity is given in Table 5. 

Reliable estimates for the transport coefficients will help us 
to understand better the creation of the primordial magnetic fields, phase
transitions in the early universe and the creation of density perturbations
in the primordial plasma, which then would seed galaxy formation.

{\begin{table*}
\centering
\caption[t5]{Summary of the values of the total conductivity $\sigma$ for 
leptons and quarks; thermal conductivity $\kappa$ and shear viscosity
$\eta$ for each particle species separately. The row 'Particles' indicates
which particles (and their antiparticles) are present in the heat bath.}
\vskip10pt
\begin{tabular}{|c|c|c|c|c|c|c|}
\hline
Particles&
e &
e $\mu$ u d &
e $\mu$ u d s &
e $\mu$ u d s c &
e $\mu \ \tau$ u d s c &
e $\mu \ \tau$ u d s c b \\
\hline
$\sigma_{tot(QED)}/T$ &
2.22 &
5.90 &
5.98 &
6.18 &
9.44 &
9.47 \\
\hline
$\sigma_{tot(QCD)}/T$ &
- &
0.110 &
0.133 &
0.223 &
0.223 &
0.247 \\
\hline
$\kappa_l/T^2$ &
4.51$\times$10$^2$ &
6.01$\times$10$^2$ &
6.07$\times$10$^2$ &
6.25$\times$10$^2$ &
6.36$\times$10$^2$ &
6.37$\times$10$^2$ \\
\hline
$\kappa_q/T^2$ &
- &
15.4 &
15.6 &
15.7 &
15.7 &
15.9 \\
\hline
$\kappa_\gamma/T^2$ &
3.58$\times$10$^4$ &
9.76$\times$10$^3$ &
8.95$\times$10$^3$ &
6.71$\times$10$^3$ &
5.65$\times$10$^3$ &
5.37$\times$10$^3$ \\
\hline
$\kappa_g/T^2$ &
- &
8.89 &
6.63 &
5.34 &
5.34 &
4.48 \\
\hline
$\kappa_{\nu_e}/T^2$ &
2.45$\times$10$^9$ &
6.63$\times$10$^8$ &
5.59$\times$10$^8$ &
3.80$\times$10$^8$ &
3.48$\times$10$^8$ &
3.28$\times$10$^8$ \\
\hline
$\kappa_{\nu_\mu}/T^2$ &
3.44$\times$10$^9$ &
6.63$\times$10$^8$ &
5.59$\times$10$^8$ &
3.80$\times$10$^8$ &
3.48$\times$10$^8$ &
3.28$\times$10$^8$ \\
\hline
$\kappa_{\nu_\tau}/T^2$ &
3.44$\times$10$^9$ &
1.72$\times$10$^9$ &
1.38$\times$10$^9$ &
1.19$\times$10$^9$ &
3.48$\times$10$^8$ &
3.28$\times$10$^8$ \\
\hline
$\eta_l/T^3$ &
4.29$\times$10$^2$ &
5.71$\times$10$^2$ &
5.76$\times$10$^2$ &
5.94$\times$10$^2$ &
6.04$\times$10$^2$ &
6.07$\times$10$^2$ \\
\hline
$\eta_q/T^3$ &
- &
17.8 &
18.0 &
18.2 &
18.2 &
18.4 \\
\hline
$\eta_\gamma/T^3$ &
2.11$\times$10$^4$ &
5.75$\times$10$^3$ &
5.27$\times$10$^3$ &
3.95$\times$10$^3$ &
3.33$\times$10$^3$ &
3.16$\times$10$^3$ \\
\hline
$\eta_g/T^3$ &
- &
4.61 &
3.50 &
2.84 &
2.84 &
2.40 \\
\hline
$\eta_{\nu_e}/T^3$ &
1.59$\times$10$^{9}$ &
4.29$\times$10$^8$ &
3.62$\times$10$^8$ &
2.46$\times$10$^8$ &
2.25$\times$10$^8$ &
2.12$\times$10$^8$ \\
\hline
$\eta_{\nu_\mu}/T^3$ &
2.22$\times$10$^{9}$ &
4.29$\times$10$^8$ &
3.62$\times$10$^8$ &
2.46$\times$10$^8$ &
2.25$\times$10$^8$ &
2.12$\times$10$^8$ \\
\hline
$\eta_{\nu_\tau}/T^3$ &
2.22$\times$10$^{9}$ &
1.12$\times$10$^{9}$ &
8.93$\times$10$^{8}$ &
7.73$\times$10$^{8}$ &
2.25$\times$10$^8$ &
2.12$\times$10$^8$ \\
\hline
\end{tabular}
\label{t5}
\end{table*}}
%\vskip30pt

\vskip10pt
{\large\bf Acknowledgements}
\vskip5pt\noindent
We wish to thank Kari Enqvist, Henning Heiselberg, Kimmo Kainulainen and 
Mika Karjalainen
for illuminating discussions on many topics presented in this paper.
%\newpage
%%%%%%%%%%%%%%%%%%%%%%%%%%%%%%%%%%%%%%%%%%%%%%%%%%%%%%%%%%%%%%%%%%%%%%%%%%%%%%

\end{document}